\documentclass[prd,aps,twocolumn,groupedaddress,showpacs]{revtex4}
\begin{document}
\input{psfig.sty}

\title{Towards a Realistic Neutron Star Binary Inspiral}

\author{ Mark Miller${}^{(1,2)}$ and Wai-Mo Suen${}^{(2,3)}$}

\address{${}^{(1)}$ 238-332 Jet Propulsion Laboratory, 4800 Oak Grove Drive,
Pasadena, CA  91109}

\address{${}^{(2)}$McDonnell Center for the Space Sciences,
Department of Physics,
Washington University, St. Louis, Missouri 63130}

\address{${}^{(3)}$Physics Department,
Chinese University of Hong Kong,
Hong Kong}

\begin{abstract}                                                        %
 
An approach to general relativity based on
conformal flatness and quasiequilibrium (CFQE) assumptions has
played an important role in the study of the inspiral
dynamics and in providing initial data for fully 
general relativistic numerical simulations of coalescing compact binaries.
However, the regime of validity of the approach has never been
established.  To this end, 
we develop an analysis that
determines the violation of the CFQE approximation in the evolution of
the binary described by the full Einstein theory.  With this analysis, we
show that
the CFQE assumption is significantly violated even at relatively large
orbital separations in the case of corotational neutron star binaries.
We also demonstrate that the innermost stable circular orbit 
(ISCO) determined in the CFQE approach for corotating neutron star binaries
may have no astrophysical significance. 

\end{abstract}


\maketitle

%
%

\paragraph {Introduction} The analysis of general relativistic binary
neutron star processes is
an important yet challenging endeavor.  The importance of studying
binary neutron star coalescence is rooted in observational astronomy;
the process is considered to be a candidate for the 
sources of detectable gravitational waves as well as $X$- and $\gamma$-rays.
Recently, there has been a large effort expended towards numerically
simulating the inspiral and coalescence of compact objects in order
to determine their signature gravitational
waveforms in preparation for the detection of gravitational waves by the
new generation of gravitational wave observatories, including LIGO,
LISA, VIRGO, TAMA, and GEO.  The complexity of the nonlinear
Einstein field equations, the lack of exact symmetry of the system, and
the coupling of different timescales in the process combine to make the
study of the neutron star inspiral coalescence a major challenge.

A recent approach used to investigate the inspiral coalescence problem that
has drawn much
attention~\cite{Baumgarte98b,Baumgarte98c,Bonazzola97,Gourgoulhon01,
Marronetti:1999ya,Shibata98,Teukolsky98,Yo01,Duez00,Duez02} is based
on the separation of two timescales, namely, the orbital motion
timescale and the gravitational radiation timescale.  This approach,
which we refer to as the conformally flat quasiequilibrium (CFQE)
approach, seeks to construct general relativistic configurations
corresponding to two compact objects in circular orbit by assuming
that the spacetime is essentially stationary with an approximate Killing
vector (quasiequilibrium, QE).  The slow secular evolution is driven
by the gravitational radiation which has a much longer timescale
than the orbital motion timescale, so that the spacetime
is thought to be describable by a 
conformally flat (CF) spatial metric.

The CFQE approach leads to a two-phase description of the inspiral
and coalescence
process.  In the ``slowly inspiraling phase'', the evolution is
approximated by a time sequence (which we refer to as a
CFQE-sequence) of CFQE
configurations with the same rest mass and spin state (e.g. corotating
or nonrotating).  Each CFQE configuration in
the time sequence is obtained by solving the four constraint equations plus
one of the dynamical Einstein equations 
under the CF and QE assumptions (the QE assumption
affects the constraint equations only indirectly).  The inspiral is
assumed to
proceed along the sequence in a quasistationary manner on a timescale
much longer than the orbital timescale.  The slow secular evolution
is driven by gravitational radiation, which can be calculated using e.g.,
the standard quadrupole formula (see, e.g.,~\cite{Wald84}).  
In the corotating case,
when the orbit shrinks to a certain radius (which will depend on the
equation of state (EOS) of the neutron stars),
the CFQE configuration becomes secularly unstable,
with the ADM mass of the system
attaining a minimum among all configurations in the sequence.  The orbit
at this radius is referred to as the innermost stable circular orbit
(ISCO).  
One expects the binary to become dynamically unstable at an orbital
radius that is slightly less than this ISCO point.  This ISCO
point is therefore thought to be the point at which
the system enters the
``plunge phase'',  in which the subsequent merging occurs within
an orbital timescale.

Besides providing a description of the inspiral coalescence process,
the CFQE approach is important for another reason: it is expected to
provide a starting point for numerical investigations in general
relativity.
The setting of initial data is a sensitive
issue in the numerical integration of the Einstein equations.
On the one hand, astrophysically relevant initial data must be used in
order to make contact with gravitational wave observations.  On the other hand, 
we are forced to use initial data in the
late state of the inspiral, due to limited
computational resources.  
As the CFQE approach gives
configurations that satisfy the constraint equations of the
Einstein theory, one is tempted to start numerical integration with a
CFQE configuration near the ISCO point as determined in the CFQE approach,
see, e.g.,~\cite{Shibata99d}. The danger 
in using the CFQE approach to
determine the initial configuration for numerical integration is the
unknown astrophysical relevance of the initial data, regardless of the
fact that each CFQE
configuration satisfies the constraints of general relativity and
is therefore, in principle, a legitimate initial data set
for numerical integrations. In order for the
results of the numerical evolutions to be relevant to observations, e.g., the
gravitational waves emitted in neutron star coalescences, we have to make sure
that the initial data actually corresponds to a configuration in a
realistic inspiral.  The setting of physical initial data is a standard
difficulty in
numerical relativity and it is particularly acute in this
case; if the initial data is picked to be a CFQE configuration in
a regime where the CFQE-sequence is {\it not} expected to be valid, the
initial data must be regarded as highly contrived and the 
resulting calculation will have no astrophysical significance.

In this paper we analyze the limitations of the CFQE approach, 
which suffers from two basic problems. The first
problem is that the CF and QE assumptions
violate the Einstein equations which, for this system, forbid
both an exact Killing vector 
and a conformally flat metric at all times (the metric can only be chosen to
be conformally flat at {\it one} point in time but not along the entire time
sequence).  The second problem is that the CFQE approach itself does
not provide a way to estimate the error involved in its assumptions.  
A determination of its accuracy has to be carried out
with a completely different analysis.  
Because of this, the regime of validity of
the CFQE approach has never been determined.  
Here, we calculate the rate of increase of the accuracy
of the CF and QE approximations with increasing initial neutron star
separation.
For the corotating binaries studied (with a
$\Gamma =2$ polytropic equation of state), 
we demonstrate that an initial separation of at least $6$ neutron star
radii is needed in order for the violation of the QE assumption
to remain below $10\%$ (in some reasonable measure defined below) after a short evolution (a fraction of an orbit).
(The violation of the CF assumption is an order of magnitude
smaller at this separation.)  It is safe to conclude that,
starting at a separation of $6$ neutron star radii,
the evolution (e.g., the orbital phase) of the system
is completely different from that of the CFQE-sequence well before
the CFQE ISCO separation (less than $3$ neutron star radii) is reached.

In what follows, we present the results of our analysis
of the CFQE-sequence approximation for binary, corotating neutron stars.
The details of all calculations and
the subsequent analysis of the numerical results,
along with results on long timescale (e.g., multiple
orbital period timescale) numerical evolutions will be
reported elsewhere~\cite{InPrepLongPaper}.

\paragraph { A General Relativistic Analysis of the CFQE Approximations for
Corotating Neutron Star Binaries.} To analyze the validity 
of the CFQE-sequence approximation, we compare it to solutions
of the full Einstein equations whose initial Cauchy slice corresponds
to specific CFQE configurations. 
We construct these corotating CFQE configurations
following the algorithm
detailed in~\cite{Baumgarte98b}.  The neutron stars we use are
described by a polytropic EOS,
$P  = K \rho^\Gamma$, with $\Gamma = 2$ and $K = 0.0445 \: {c^2}/{\rho_n}$,
where $\rho_n$ is nuclear density
($2.3$  x  $10^{14} \; g/cm^3$). 
For this EOS, the maximum stable static neutron star configuration
has an ADM mass of $1.79 M_{\odot}$ and a baryonic mass
of $1.97 M_{\odot}$ ($M_\odot$ is 1 solar mass). 
In this paper, we use neutron stars whose baryonic mass $M_0$ is 
$1.49 \: M_\odot$, which
is approximately $75\%$ that
of the maximum stable configuration.  The ADM mass of a single static
neutron star for this configuration
is $1.4 M_\odot$.

The corotational CFQE configurations with a fixed baryonic mass can be uniquely
specified by the separation between the two neutron stars.   We measure the
spatial separation on a constant time hypersurface by the
spatial geodesic distance $\ell_{1,2}$ between the points of maximum
baryonic mass density in each of the neutron stars.  This 
particular choice of measure of
distance depends on the choice of time slicing.  In order
to compare with the CFQE-sequence approximation in an invariant manner,
we use the same slicing condition (maximal slicing) 
as that used in CFQE-sequence approach.  Four CFQE configurations denoted respectively NS-1 to NS-4 are used as initial data in
our general relativistic simulations in this paper: $\ell_{1,2}/ M_0=23.44, 25.94, 29.78, 35.72$.  The corresponding orbital angular velocity parameters are $\Omega M_0 = 0.01547, 0.01296, 0.01022, 0.00746$.

In the following we first take care to construct invariant measures that can be used to
compare, and give a sense of the difference between, two different
spacetimes, one of which is obtained by solving the Einstein equations and the
other represented by the CFQE-sequence, which does not satisfy the
Einstein equations.

\paragraph {The QE approximation.}  One basic assumption of
the CFQE approach is the existence of a timelike, helical Killing
vector field.  In the case of a corotating binary, the
4-velocity $u^a$  of the fluid is proportional to this Killing vector.
This implies the vanishing of the type-(0,2) 4-tensor $Q_{ab} \equiv \nabla_a u_b + \nabla_b u_a + u_a a_b + u_b a_a ,$
where $a^a \equiv u^b \nabla_b u^a$ is the 4-acceleration of the
fluid ($\nabla_a$ denotes the covariant derivative operator
compatible with the 4-metric $g_{ab}$).
The quantity $Q_{ab} u^a$ vanishes identically.
We can thus monitor the space-space components of $Q_{ab}$ during
full general relativistic simulations as a way of monitoring how
well the 4-velocity $u^a$ stays proportional to a
Killing vector field.  Define $Q_{ij}$ as the projection
of $Q_{ab}$ onto the constant $t$ spatial slice: $Q_{ij} = {P_i}^a {P_j}^b Q_{ab} =
   Q_{ab} {\left ( \frac{\partial}{\partial x^i} \right )}^a
          {\left ( \frac{\partial}{\partial x^j} \right )}^b ,$
where $P_{ab} = g_{ab} + n_a n_b$ is the projection operator
onto the constant $t$ spatial slices with unit normal $n^a$.  The 3-coordinate independent norm $\left | Q_{ij} \right | $ of this matrix $Q_{ij}$ is the square root of the
largest eigenvalue of $Q_{ij} {Q^j}_k$, where we have raised
and lowered 3-indices with the 3-metric.  

To gain a sense of the size of
$\left | Q_{ij} \right | $, we normalize $\left | Q_{ij} \right | $
by the norms of its two principle parts: $ Q \equiv {\left | Q_{ij} \right |}/
          {\mbox{max} \{ \left | {Q_1}_{ij} \right | ,
                         \left | {Q_2}_{ij} \right | \} },$
where ${Q_1}_{ab} \equiv \nabla_a u_b + \nabla_b u_a$, and ${Q_2}_{ab}  \equiv u_a a_b + u_b a_a. $
A value of $Q=0$ signifies that the 4-velocity of the fluid
is proportional to a timelike Killing vector as in the CFQE-sequence,
while a value of $Q=1$ would signify that the assumption is maximally
violated.

Since $Q$ is 
meaningful only inside the fluid bodies, a natural global measure of the magnitude of $Q$ is its the baryonic mass weighted integral, denoted by
$\langle Q \rangle$:
\begin{equation}
\label{eq:weight_q}
\langle Q \rangle =
   \frac {\int d^3\!x \: \left | Q \right | \sqrt{\gamma} \rho W }
   {\int d^3\!x \: \sqrt{\gamma} \rho W},
\end{equation}
where the integrals are taken to be over the entire spatial slice, with
support only where $\rho \neq 0$.  Here, we have denoted $\gamma$ to be
the determinant of the 3-metric and $W$ to be the Lorentz factor
of the fluid so that ${\int d^3\!x \: \sqrt{\gamma} \rho W}$ is the
total (conserved) rest mass of the system.  

\begin{figure}
\vspace{0.0cm}
\hspace{0.0cm}
\psfig{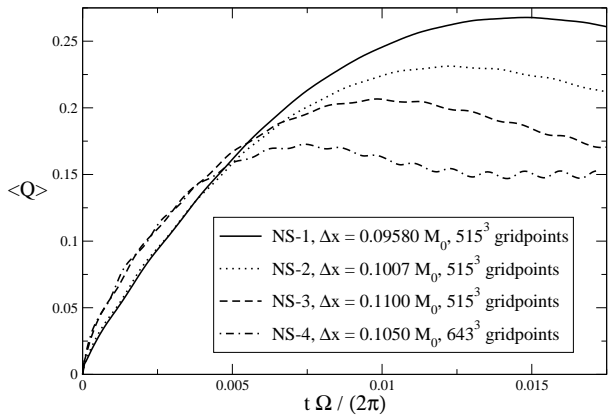}
\caption{
The quantity $\langle Q \rangle$ is plotted as a
function
of time
for fully consistent general relativistic numerical
calculations, using CFQE configuration NS-1 to NS-4 as initial data.  
}
\vspace{0.0cm}
\label{fig:q_ns1234}
\end{figure}

In Fig.~\ref{fig:q_ns1234}, we plot $\langle Q \rangle$ as
a function of time in the general
relativistic numerical simulations, using CFQE configurations
NS-1 to NS-4 as initial data.
For each initial CFQE configuration, the subsequent solution
to the Einstein equations has $\langle Q \rangle$ reaching a maximum
value after a very short time (between $0.5\%$ and $1.5\%$ of an orbit,
$2 \pi / \Omega$ being the orbital period corresponding to 
each CFQE configuration).  
These results are in contrast to the CFQE-sequence
approximation, where $\langle Q \rangle$ is exactly zero.
The rapid growth of $\langle Q \rangle$ 
to such a large value 
signals that, in the present case of corotating neutron stars, after just a short time
the CFQE-sequence is very
different from the spacetime described by the full Einstein equations
even when the latter is started with the same initial data.   Within a
small fraction of an orbit, the 4-velocity of the fluid already is in
no way proportional to a Killing vector.

Also, we see from Fig.~\ref{fig:q_ns1234}
that the maximum value obtained by $\langle Q \rangle$ decreases
from approximately $\langle Q \rangle = 0.27$ to
$\langle Q \rangle = 0.17$ as the geodesic separation
of the initial CFQE configuration increases from
$\ell_{1/2}/M_0 = 23.44$ to $\ell_{1/2}/M_0 = 35.72$.
In principle one can obtain an arbitrarily accurate CFQE configuration by
starting the inspiral evolution with a large enough separation.  While
starting at a larger separation does {\it not} imply the CFQE-sequence will be
more astrophysically relevant at 
late times, the fully relativistic simulation starting
with it will be. 
This effect can be seen directly in
Fig.~2a, where the maximum value of $\langle Q \rangle$
obtained
in the general relativistic simulations is
plotted
as a function of the initial geodesic separation $\ell_{1,2}$. 
A careful analysis of the numerical errors
(shown as error bars in the figure)
based on a Richardson extrapolation technique
(for {\it both} truncation errors and boundary errors)
can be found in~\cite{InPrepLongPaper}.
\begin{figure}
\vspace{0.0cm}
\hspace{0.0cm}
\psfig{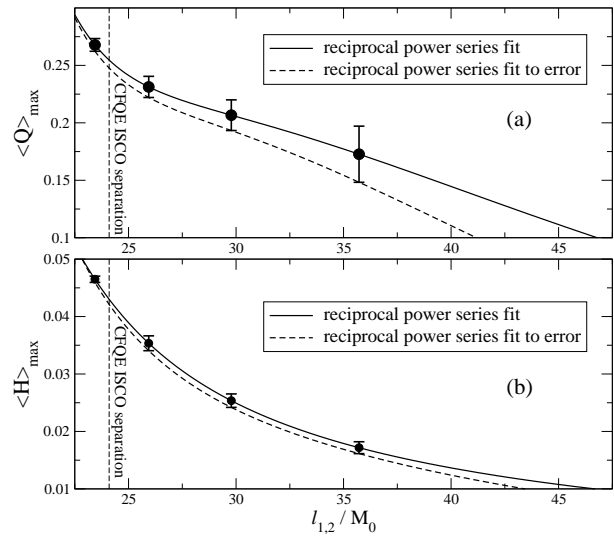}
\caption{
Fig.~2a shows the maximum value of $\langle Q \rangle$ 
obtained in
our fully consistent general relativistic simulations using CFQE
configurations with different geodesic separation $\ell_{1,2}$ as
initial data.  Fig.~2b shows the corresponding 
maximum value of $\langle H \rangle$ 
}
\vspace{0.0cm}
\label{fig:maxq}
\end{figure}
In this way, we can
determine the required separation 
of the initial CFQE configuration such that the QE assumption is
valid to any required level in our fully consistent general relativistic
simulations. In
Fig.~2a, we fit a reciprocal power law,
$\frac {a_1}{{(\ell_{1,2})}} + \frac {a_2}{{(\ell_{1,2})}^2} +
\frac {a_3}{{(\ell_{1,2})}^3} + \frac {a_4}{{(\ell_{1,2})}^4}$,
to the maximum value attained in our
general relativistic calculations (as well as to the lower
bound of the estimated error).  We can see that, e.g.,
one would have to use a CFQE configuration initial data set with
a geodesic separation between the neutron stars of
about $\ell_{1,2} = 47 \: M_0$ in order for the subsequent solution to
have the fluid 4-velocity following a Killing vector to the $10\%$ level.

\paragraph{The CF Approximation.}

\begin{figure}
\vspace{0.0cm}
\hspace{0.0cm}
\psfig{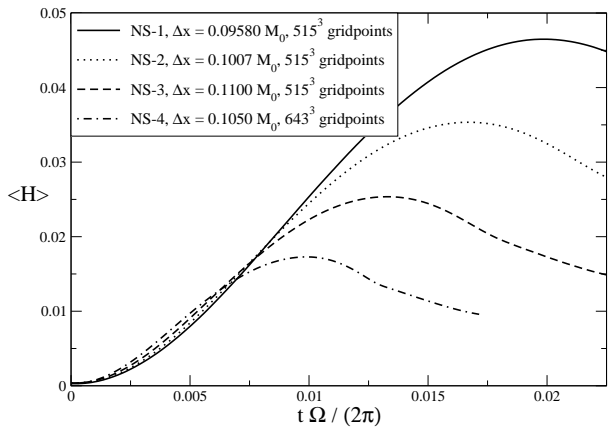}
\caption{
The quantity $\langle H \rangle$ is plotted as a
function
of time
for fully consistent general relativistic numerical
calculations, using CFQE configuration NS-1 to NS-4 as initial data.
}
\vspace{0.0cm}
\label{fig:hrho_zap1234}
\end{figure}

While the 3-metric can be chosen to be
conformally flat on any one time slice,  the assumption that the
3-metric remain conformally flat for all time violates the dynamical
Einstein equations.  Again we begin with an
invariant construction of a measure of
the violation.   Because conformal flatness 
is a property of a constant time slice, and because we are
using the same slicing conditions in our simulation as that
used in the CFQE-sequence approximation,
we seek a 3-invariant (with respect to
spatial coordinate transformations) to monitor the CF assumption.
The Bach 3-tensor $B_{ijk} = 2 {\cal D}_{[i} \left ( {}^{(3)}\!R_{j]k} -
   \frac {1}{4} {\gamma}_{j]k} {}^{(3)}\!R \right )$
can be shown to vanish if and only if the 3-metric
$\gamma_{ij}$ is conformally flat.  The Cotton-York tensor,
$H_{ij}$, is related to the 3-Bach tensor by $H_{ij} = {{\epsilon}^{mn}}_j B_{mni}, $
where ${\epsilon}_{ijk}$ is the natural volume element 3-form.
We define the scalar $H$ as the matrix norm of $H_{ij}$, 
normalized by the size of the covariant derivative
of the 3-Ricci tensor: $ H = {\left | H_{mn} \right |} /
          {\sqrt{{\cal D}_i {}^{(3)}\!R_{jk}
                 {\cal D}^i {}^{(3)}\!R^{jk}}},$
where ${\left | H_{ij} \right |}$ denotes
the matrix norm of the $H_{ij}$.
As before, we define
the baryonic mass density weight of H, denoted
as $\langle H \rangle$, by
\begin{equation}
\langle H \rangle =
   \frac {\int d^3\!x \: \left | H \right | \sqrt{\gamma} \rho W }
   {\int d^3\!x \: \sqrt{\gamma} \rho W },
\label{eq:weight_h_rho}
\end{equation}
where the integrals are taken to be over the entire spatial slice,
but have support only where $\rho \neq 0$.

In Fig.~\ref{fig:hrho_zap1234}, we plot $\langle H \rangle$ as a
function
of time for the general relativistic simulations using initial
data CFQE configurations NS-1 through NS-4.  The profile of the 
violation of the conformal flatness assumption $\langle H \rangle$
is similar to that of the Killing vector assumption $\langle Q \rangle$:
a quick rise to a maximum value at a timescale on the 
same order of but slightly longer than that of $\langle Q \rangle$.
This is contrasted to the CFQE-sequence
approximation, which has $\langle H \rangle$ exactly equal to $0$.

In Fig.~2b, we plot the maximum value of the
quantity $\langle H \rangle$ as a function of initial geodesic
separation
$\ell_{1,2}$ attained in our general
relativistic numerical simulations using the four CFQE configurations
NS-1 through NS-4 as initial data (see~\cite{InPrepLongPaper}
for a detailed analysis of the truncation and boundary errors in our
calculation, which we indicate as error bars in the figure).
As in Fig.~2a, we fit a reciprocal power law to
both the maximum value of $\langle H \rangle$ and the lower bound 
of the error.  We see that a CFQE configuration with
geodesic separation $\ell_{1,2}$ of approximately $47 \: M_0$ or
greater must be used as initial data in
order for the full solution to the Einstein field equations 
to have a
value of $\langle H \rangle$ to be
$0.01$ or less.

\paragraph{Conclusions}

In this letter we present a method for analyzing the regime of
validity of the CFQE-sequence approximation.  We apply this method 
to the CFQE-sequence approximation of corotating binary neutron stars.
We show that for geodesic separations 
less than roughly $45 \: M_0$, which is about twice the geodesic 
separation of the ISCO configuration, 
the QE approximation is severely violated and the CFQE
sequence cannot be taken as a reasonable approximation of the binary
evolution for these separations.  
Numerical simulations starting with CFQE configurations as initial
data with geodesic separations smaller than $45 \: M_0$ cannot,
therefore, be considered as approximating a realistic neutron star
binary inspiral.

\section{Acknowledgement}
\label{ack}

It is a pleasure to thank
Abhay Ashtekar, Lee Lindblom, Kip Thorne,
and Clifford Will
for useful discussions and comments.
Our application code, which solves the coupled Einstein-relativistic 
hydrodynamics equations uses the Cactus Computational 
Toolkit~\cite{Cactusweb} for parallelization and high performance I/O.

Financial support for this research has been
provided by the
NSF KDI Astrophysics Simulation
Collaboratory (ASC) project (Phy 99-79985) and the
Jet Propulsion Laboratory (account 100581-A.C.02).
Computational resource support has been provided by the
NSF NRAC grants MCA02N022 and MCA93S025.


\begin{thebibliography}{13}
\expandafter\ifx\csname natexlab\endcsname\relax\def\natexlab#1{#1}\fi
\expandafter\ifx\csname bibnamefont\endcsname\relax
  \def\bibnamefont#1{#1}\fi
\expandafter\ifx\csname bibfnamefont\endcsname\relax
  \def\bibfnamefont#1{#1}\fi
\expandafter\ifx\csname citenamefont\endcsname\relax
  \def\citenamefont#1{#1}\fi
\expandafter\ifx\csname url\endcsname\relax
  \def\url#1{\texttt{#1}}\fi
\expandafter\ifx\csname urlprefix\endcsname\relax\def\urlprefix{URL }\fi
\providecommand{\bibinfo}[2]{#2}
\providecommand{\eprint}[2][]{\url{#2}}

\bibitem[{\citenamefont{Baumgarte
  et~al.}(1998{\natexlab{a}})\citenamefont{Baumgarte, Cook, Scheel, Shapiro,
  and Teukolsky}}]{Baumgarte98b}
\bibinfo{author}{\bibfnamefont{T.~W.} \bibnamefont{Baumgarte}},
  \bibinfo{author}{\bibfnamefont{G.~B.} \bibnamefont{Cook}},
  \bibinfo{author}{\bibfnamefont{M.~A.} \bibnamefont{Scheel}},
  \bibinfo{author}{\bibfnamefont{S.~L.} \bibnamefont{Shapiro}},
  \bibnamefont{and} \bibinfo{author}{\bibfnamefont{S.~A.}
  \bibnamefont{Teukolsky}}, \bibinfo{journal}{Physical Review D}
  \textbf{\bibinfo{volume}{57}}, \bibinfo{pages}{7299}
  (\bibinfo{year}{1998}{\natexlab{a}}).

\bibitem[{\citenamefont{Baumgarte
  et~al.}(1998{\natexlab{b}})\citenamefont{Baumgarte, Cook, Scheel, Shapiro,
  and Teukolsky}}]{Baumgarte98c}
\bibinfo{author}{\bibfnamefont{T.~W.} \bibnamefont{Baumgarte}},
  \bibinfo{author}{\bibfnamefont{G.~B.} \bibnamefont{Cook}},
  \bibinfo{author}{\bibfnamefont{M.~A.} \bibnamefont{Scheel}},
  \bibinfo{author}{\bibfnamefont{S.~L.} \bibnamefont{Shapiro}},
  \bibnamefont{and} \bibinfo{author}{\bibfnamefont{S.~A.}
  \bibnamefont{Teukolsky}}, \bibinfo{journal}{Physical Review D}
  \textbf{\bibinfo{volume}{57}}, \bibinfo{pages}{6181}
  (\bibinfo{year}{1998}{\natexlab{b}}).

\bibitem[{\citenamefont{Bonazzola et~al.}(1997)\citenamefont{Bonazzola,
  Gourgoulhon, and Marck}}]{Bonazzola97}
\bibinfo{author}{\bibfnamefont{S.}~\bibnamefont{Bonazzola}},
  \bibinfo{author}{\bibfnamefont{E.}~\bibnamefont{Gourgoulhon}},
  \bibnamefont{and} \bibinfo{author}{\bibfnamefont{J.-A.} \bibnamefont{Marck}},
  \bibinfo{journal}{Phys. Rev. D} \textbf{\bibinfo{volume}{56}},
  \bibinfo{pages}{7740} (\bibinfo{year}{1997}).

\bibitem[{\citenamefont{Gourgoulhon et~al.}(2001)\citenamefont{Gourgoulhon,
  Grandclement, Taniguchi, Marck, and Bonazzola}}]{Gourgoulhon01}
\bibinfo{author}{\bibfnamefont{E.}~\bibnamefont{Gourgoulhon}},
  \bibinfo{author}{\bibfnamefont{P.}~\bibnamefont{Grandclement}},
  \bibinfo{author}{\bibfnamefont{K.}~\bibnamefont{Taniguchi}},
  \bibinfo{author}{\bibfnamefont{J.}~\bibnamefont{Marck}}, \bibnamefont{and}
  \bibinfo{author}{\bibfnamefont{S.}~\bibnamefont{Bonazzola}},
  \bibinfo{journal}{Phys. Rev. D} \textbf{\bibinfo{volume}{63}},
  \bibinfo{pages}{064029} (\bibinfo{year}{2001}).

\bibitem[{\citenamefont{Marronetti et~al.}(1999)\citenamefont{Marronetti,
  Mathews, and Wilson}}]{Marronetti:1999ya}
\bibinfo{author}{\bibfnamefont{P.}~\bibnamefont{Marronetti}},
  \bibinfo{author}{\bibfnamefont{G.~J.} \bibnamefont{Mathews}},
  \bibnamefont{and} \bibinfo{author}{\bibfnamefont{J.~R.}
  \bibnamefont{Wilson}}, \bibinfo{journal}{Phys. Rev.}
  \textbf{\bibinfo{volume}{D60}}, \bibinfo{pages}{087301}
  (\bibinfo{year}{1999}).

\bibitem[{\citenamefont{Shibata}(1998)}]{Shibata98}
\bibinfo{author}{\bibfnamefont{M.}~\bibnamefont{Shibata}},
  \bibinfo{journal}{Phys. Rev. D} \textbf{\bibinfo{volume}{58}},
  \bibinfo{pages}{024012} (\bibinfo{year}{1998}).

\bibitem[{\citenamefont{Teukolsky}(1998)}]{Teukolsky98}
\bibinfo{author}{\bibfnamefont{S.}~\bibnamefont{Teukolsky}},
  \bibinfo{journal}{ApJ} \textbf{\bibinfo{volume}{504}}, \bibinfo{pages}{442}
  (\bibinfo{year}{1998}).

\bibitem[{\citenamefont{Yo et~al.}(2001)\citenamefont{Yo, Baumgarte, and
  Shapiro}}]{Yo01}
\bibinfo{author}{\bibfnamefont{H.-J.} \bibnamefont{Yo}},
  \bibinfo{author}{\bibfnamefont{T.}~\bibnamefont{Baumgarte}},
  \bibnamefont{and} \bibinfo{author}{\bibfnamefont{S.}~\bibnamefont{Shapiro}},
  \bibinfo{journal}{Phys. Rev. D} \textbf{\bibinfo{volume}{63}}
  (\bibinfo{year}{2001}).

\bibitem[{\citenamefont{Duez et~al.}(2001)\citenamefont{Duez, Baumgarte, and
  Shapiro}}]{Duez00}
\bibinfo{author}{\bibfnamefont{M.~D.} \bibnamefont{Duez}},
  \bibinfo{author}{\bibfnamefont{T.~W.} \bibnamefont{Baumgarte}},
  \bibnamefont{and} \bibinfo{author}{\bibfnamefont{S.~L.}
  \bibnamefont{Shapiro}}, \bibinfo{journal}{Phys. Rev.}
  \textbf{\bibinfo{volume}{D63}}, \bibinfo{pages}{084030}
  (\bibinfo{year}{2001}).

\bibitem[{\citenamefont{Duez et~al.}(2002)\citenamefont{Duez, Baumgarte,
  Shapiro, Shibata, and Uryu}}]{Duez02}
\bibinfo{author}{\bibfnamefont{M.~D.} \bibnamefont{Duez}},
  \bibinfo{author}{\bibfnamefont{T.~W.} \bibnamefont{Baumgarte}},
  \bibinfo{author}{\bibfnamefont{S.~L.} \bibnamefont{Shapiro}},
  \bibinfo{author}{\bibfnamefont{M.}~\bibnamefont{Shibata}}, \bibnamefont{and}
  \bibinfo{author}{\bibfnamefont{K.}~\bibnamefont{Uryu}},
  \bibinfo{journal}{Phys. Rev.} \textbf{\bibinfo{volume}{D65}},
  \bibinfo{pages}{024016} (\bibinfo{year}{2002}).

\bibitem[{\citenamefont{Wald}(1984)}]{Wald84}
\bibinfo{author}{\bibfnamefont{R.~M.} \bibnamefont{Wald}},
  \emph{\bibinfo{title}{General Relativity}} (\bibinfo{publisher}{The
  University of Chicago Press}, \bibinfo{address}{Chicago},
  \bibinfo{year}{1984}).

\bibitem[{\citenamefont{Shibata and Uryu}(2000)}]{Shibata99d}
\bibinfo{author}{\bibfnamefont{M.}~\bibnamefont{Shibata}} \bibnamefont{and}
  \bibinfo{author}{\bibfnamefont{K.}~\bibnamefont{Uryu}},
  \bibinfo{journal}{Phys. Rev. D} \textbf{\bibinfo{volume}{61}},
  \bibinfo{pages}{064001} (\bibinfo{year}{2000}).

\bibitem{InPrepLongPaper}
\bibinfo{author}{\bibfnamefont{M.} \bibnamefont{Miller}},
  \bibinfo{author}{\bibfnamefont{P.} \bibnamefont{Gressman}}, \bibnamefont{and}
  \bibinfo{author}{\bibfnamefont{W.}~\bibnamefont{Suen}},
  \emph{\bibinfo{title}{in preparation}}.

\bibitem[{\citenamefont{Cactus}()}]{Cactusweb}
\bibinfo{author}{\bibnamefont{Cactus}},
  \emph{\bibinfo{title}{http://www.cactuscode.org}}.

\end{thebibliography}


\end{document}